\documentclass[twocolumn,showpacs,preprintnumbers,amsmath,amssymb]{revtex4}

\usepackage{graphicx}
\usepackage{dcolumn}
\usepackage{bm}

\begin{document}
\title{Nature of electron Zitterbewegung in crystalline solids}
\date{\today}
\author{Wlodek Zawadzki*} \author{Tomasz M. Rusin \dag}
\email{zawad@ifpan.edu.pl}

\affiliation{* Institute of Physics, Polish Academy of Sciences, Al. Lotnik\'ow 32/46, 02-688 Warsaw, Poland;  \\
          \dag PTK Centertel Sp. z o.o., ul. Skierniewicka 10A, 01-230 Warsaw, Poland }


\begin{abstract}
We demonstrate both classically and quantum mechanically that the Zitterbewegung (ZB, the trembling motion)
of electrons in crystalline solids is nothing else, but oscillations of velocity assuring the
energy conservation when the electron
moves in a periodic potential. This means that the nature of electron ZB in a solid is completely
different from that of relativistic electrons in a vacuum, as proposed by Schrodinger. Still, we show that
the two-band {\bf k.p} model of electronic band structure, formally similar to the Dirac equation
for electrons in a vacuum, gives a very good description of ZB in solids. Our results indicate
unambiguously that the trembling motion of electrons in solids should be observable.
\end{abstract}

\maketitle
The phenomenon of electron Zitterbewegung (ZB, the trembling motion) was predicted by Schrodinger in 1930 as
a consequence of the Dirac equation for a free relativistic electron \cite{Schroedinger30}.
Schrodinger observed that the electron velocity components, given in the Dirac formalism by
$4 \times 4$ number operators, do not commute with the Dirac Hamiltonian, so the electron
velocity in a vacuum is not a constant of the motion also in absence of
external fields. The observability of ZB for electrons in a vacuum was debated ever since
(see e.g. \cite{Huang52,Krekora2004}).
Experimental difficulties to observe the ZB in a vacuum are great because the predicted frequency
of the trembling is very high: $\hbar \omega_Z \simeq 2m_0c^2\simeq  $ 1MeV, and its amplitude is very small:
$\lambda_c=\hbar/m_0c=3.86 \times 10^{-3}$\AA. It was later suggested that a phenomenon analogous to
ZB should exist for electrons in semiconductors if they can be described by a two-band model of
band structure \cite{Cannata90,Vonsovsky93}.
In particular, an analogy between the behavior of free relativistic electrons
in a vacuum and that of non-relativistic electrons in narrow gap semiconductors (NGS) was
used to predict that the ZB of electrons in NGS should have the
frequency $\hbar \omega_Z \simeq E_g$ (where $E_g$ is the energy gap),
and the amplitude $\lambda_Z=\hbar/m^*u$, where $m^*$ is the effective mass
and $u=(E_g/2m^*)^{1/2} \simeq 10^8$cm/s is a maximum electron velocity according to the two-band
{\bf k.p} model \cite{ZawadzkiHMF}.
This results in much more advantageous characteristics of ZB as compared to a
vacuum; in particular $\lambda_Z \simeq 64$\AA\ for InSb, $37$\AA\ for InAs, and $13$\AA\ for GaAs.
After the papers of Zawadzki \cite{Zawadzki05KP} and Schliemann {\it et al.} \cite{Schliemann05}
the ZB of electrons in crystalline solids
and other systems became a subject of intensive theoretical studies, see \cite{Rusin09}.
A classical phenomenon analogous to the ZB
was recently observed in macroscopic sonic crystals \cite{Zhang08b}.

The physical origin of ZB remained mysterious. As to electrons in a vacuum, it was
recognized that, being of the quantum nature, the phenomenon goes beyond Newton's First Law.
Also, it was remarked that the ZB is due to an interference of states corresponding to positive
and negative electron energies \cite{BjorkenBook}.
Since the ZB in solids was treated by a two-band  Hamiltonian similar to the Dirac equation,
its interpretation was also similar. This did not explain its origin, it only provided a way to describe it.
However, it was clear that, since the energy bands result from electron motion
in a crystalline periodic potential, in the final count it is this
potential that is responsible for the ZB.  Our paper treats the fundamentals of electron
propagation in a periodic potential and elucidates the nature of Zitterbewegung in solids.
The physical origin of ZB is of great importance because
it resolves the essential question of its observability.
The second purpose of our work is to decide whether the two-band {\bf k.p} model of the band structure, used
until now to describe the ZB in solids, is adequate.

It is often stated that an electron moving in a periodic potential behaves like a
free particle characterized by an effective mass $m^*$. The above picture suggests that, if there
are no external forces, the electron moves in a crystal with a constant velocity. This, however, is
clearly untrue because the electron velocity operator $\hat{v}_i=\hat{p}_i/m_0$ does
not commute with the Hamiltonian $\hat{H}=\hat{\bm p}^2/2m_0 + V(\bm r)$, so that $\hat{v}_i$ is not a
constant of the motion. In reality, as the electron moves in a periodic potential, it
accelerates or slows down keeping its total energy constant. This situation is analogous to that
of a roller-coaster: as it goes down losing its potential energy, its velocity
(i.e. its kinetic energy) increases, and when it goes up its velocity decreases.
We demonstrate below that {\it the electron velocity
oscillations due to its motion in a periodic potential of a solid} are {\it in fact the Zitterbewegung.}
Thus the electron Zitterbewegung in solids is not an exotic obscure phenomenon $-$ in reality it
describes the basic electron propagation in periodic potentials.

The first argument relates to the trembling frequency $\omega_Z$. The latter is easy to determine
if we assume, in the first approximation, that the electron moves with a constant average velocity
$\bar{v}$ and the period of the potential is $a$, so $\omega_Z=2\pi \bar{v}/a$.
Putting typical values for GaAs: $a=6.4$\AA, $\bar{v}=2.5 \times 10^7$cm/s,
one obtains $\hbar \omega_Z=1.62$eV, i.e. the interband frequency corresponding to
the energy gap $E_g \simeq 1.5$eV. The interband frequency is in fact typical for the ZB in solids.
Next we describe the velocity oscillations classically, assuming for simplicity a one-dimensional
periodic potential of the form $V(z)=V_0\sin(2\pi z/a)$. The first integral of the motion expressing
the total energy is: $E=m_0v_z^2/2 + V(z)$. Thus the velocity is
\begin{equation}
\frac{dz}{dt} = \sqrt{\frac{2E}{m_0}}\left[1-\frac{V(z)}{E} \right]^{1/2}.
\end{equation}
One can now separate the variables and integrate each side in the standard way. However, trying
to obtain an analytical result we assume $V(z) \simeq E/2$, expand the square root retaining
first two terms and solve the remaining equation by iteration taking in the  first step
a constant velocity $v_{z0}=(2E/m_0)^{1/2}$.
This gives $z=v_{z0}t$ and
\begin{equation}
v_z(t) = v_{z0}- \frac{v_{z0}V_0}{2E}\sin\left(\frac{2\pi v_{z0}t}{a}\right).
\end{equation}
Thus, as a result of the motion in a periodic potential the electron velocity oscillates with the
expected frequency $\omega_z=2\pi v_{z0}/a$ around the average value $v_{z0}$. Integrating
with respect to time to get an amplitude of ZB we obtain $\Delta z = V_0a/(4\pi E)$.
Taking again $V_0\simeq E/2$, and estimating the lattice constant to be $a\simeq \hbar p_{cv}/(m_0E_g)$
(see Luttinger and Kohn \cite{Luttinger55}), we have finally $\Delta z \simeq \hbar p_{cv}/(8\pi m_0E_g)$,
where $p_{cv}$ is the interband matrix element of momentum. This should be compared with an
estimation obtained previously from the two-band  {\bf k.p}
model \cite{Zawadzki05KP}: $\Delta z \simeq \lambda_Z = \hbar/m^*u=
\hbar(2/m^*E_g)^{1/2} \simeq 2\hbar p_{cv}/m_0E_g$. Thus the classical and
quantum results depend the same way on the fundamental parameters, although the classical
approach makes no use of the energy band structure. We conclude that the Zitterbewegung in solids
is simply due to the electron velocity oscillations assuring the energy conservation during
motion in a periodic potential.

Now we describe the electron velocity oscillations using a quantum approach. We begin with the
periodic Hamiltonian $\hat{H}=\hat{p}^2/2m_0 + V(z)$. The velocity operator is $\hat{v}_z=\hat{p}_z/m_0$.
Using the above Hamiltonian one obtains
\begin{equation} \label{PartialVz}
m_0\frac{d\hat{v}_z}{dt} = \frac{1}{i\hbar}[\hat{p}_z,\hat{H}] = -\frac{\partial V(z)}{\partial z},
\end{equation}
which is a quantum analogue of the Newton law of motion in an operator form. In order to integrate
Eq. (\ref{PartialVz}) we assume a particularly simple periodic saw-like potential. It is
described by $V(z)=-gz$ for $0\leq z  < a/2$,  $a\leq z < 3a/2$, etc., and
$V(z)=-V_0+gz$ for $a/2\leq z < a$, $(3a/2)\leq z \leq 2a$, etc., where $g$ is a constant force, see Fig 1a.
In each half-period $z$ is counted from zero. The derivatives are
$-\partial V/\partial z = g$ for $0\leq z  < a/2$,  $a\leq z < 3a/2$, etc., and
$-\partial V/\partial z = -g$ for $a/2\leq z  < a$, $(3a/2)\leq z \leq 2a$, etc.,
as illustrated in Fig. 1b. Thus the electron
moves initially with a constant acceleration $g/m_0$ from $z=0$ to $z=a/2$, reaches the maximum
velocity $\hat{v}_m=(ag/m_0)^{1/2}$ at a time $t_m= (m_0a/g)^{1/2}$, and then slows down
reaching $\hat{v}=0$ at a time $2t_m$. Then the cycle is periodically repeated. We calculate
$\hat{v}(t)=(g/m_0)t$ and $\hat{z}(t)=(g/2m_0)t^2$ in the first, third, fifth time (or distance)
intervals, and $\hat{v}(t) = v_m-(g/m_0)t$ and $\hat{z}(t)=v_mt - (g/2m_0)t^2$ in the second,
fourth, sixth time (or distance) intervals, as illustrated in Figs. 1c and 1d. We assumed for
simplicity $\hat{v}(0)=\hat{z}(0)=0$.

\begin{figure}
\includegraphics[width=8.5cm,height=8.5cm]{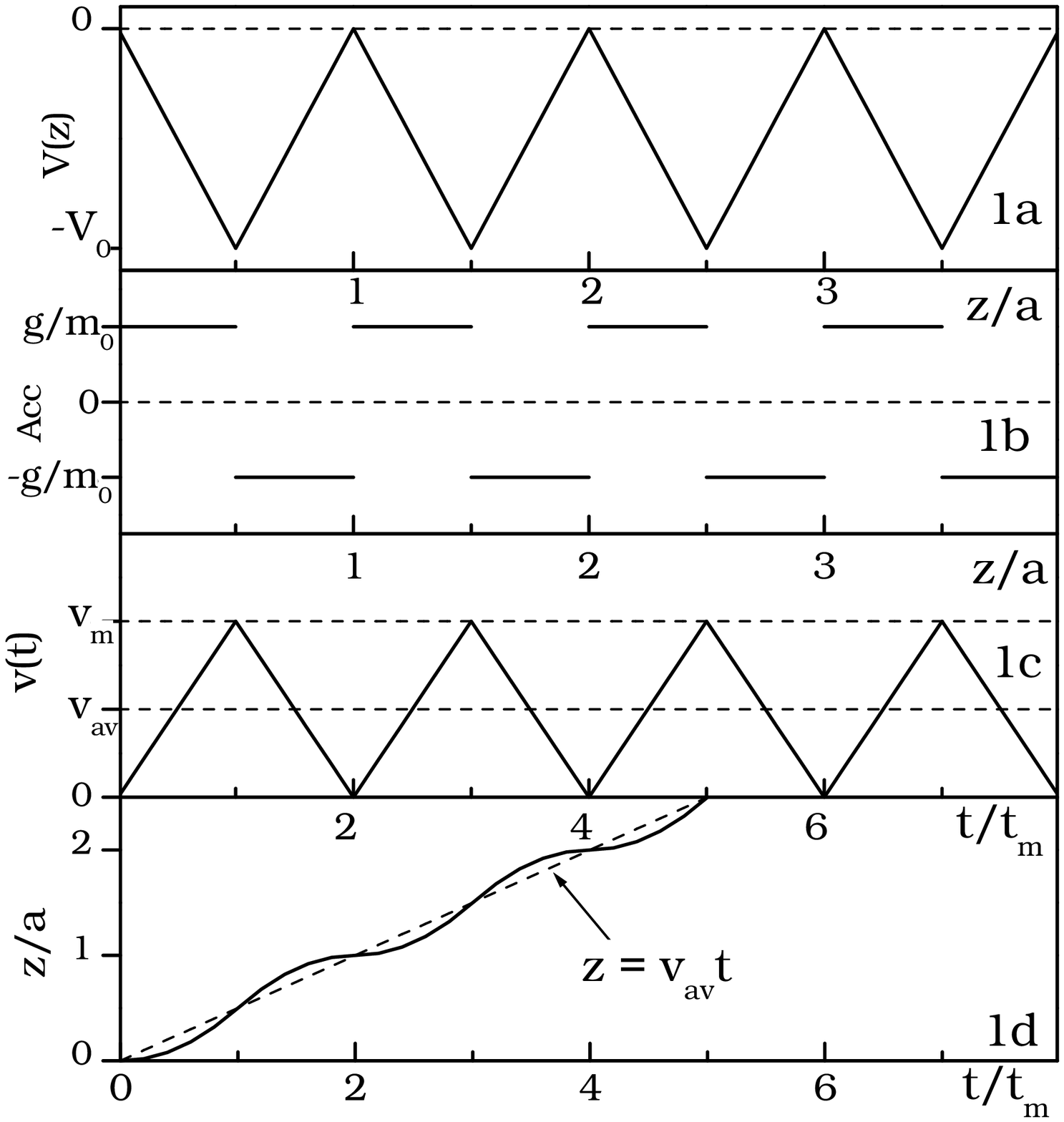}
\caption{Plot of (a) potential, (b) acceleration,  (c) velocity,  and d) position for the
         electron  motion in a saw-like potential (schematically).}
\label{Fig1}
\end{figure}

It is seen from Fig. 1c that the velocity oscillates in time around the average value $v_{av}=v_m/2$. This is
reflected in the oscillations of position $\Delta z(t)$ around the straight line $z=v_{av}t$, as seen in Fig. 1d.
The latter are easily shown to be $\Delta z= \pm a/8$, which compares well with the classical results.
Using the above estimation for $a$ we identify again the motion due to
periodicity of the potential with the Zitterbewegung calculated previously with the use of two-band
{\bf k.p} model.

Strictly speaking, the physical sense of the above operator reasoning is reached when one calculates average values.
Our procedure follows the original approach of Schrodinger \cite{Schroedinger30}, who integrated operator equations
for $\hat{v}(t)$ and $\hat{z}(t)$. Similar approach is commonly used for
$\hat{z}(t)$ and $\hat{p}(t)$ operators in the harmonic oscillator problem \cite{MessiahBook}.
Equation of motion (\ref{PartialVz})
does not contain the total electron energy but it is clear from Fig. 1a that, in a quantum treatment,
the electron can move either in allowed energy bands below the potential tops at
$E\leq 0$ or above the tops at $E>0$.

Now we describe the ZB using a rigorous quantum approach. We employ the Kronig-Penney
delta-like potential since it allows us to calculate explicitly the eigenenergies and eigenfunctions
 \cite{Kronig31, SmithBook}.
In the extended zone scheme the Bloch function is $\psi_k(z)=e^{ikz}A_k(z)$, where
\begin{equation} \label{Kron_Ak}
A_k(z) = e^{-ikz}C_k\left\{e^{ika}\sin[\beta_kz] + \sin[\beta_k(a-z)]\right\},
\end{equation}
in which $k$ is the wave vector, $C_k$ is a normalizing constant and $\beta_k = \sqrt{2m_0E}/\hbar$ is a
solution of the equation
\begin{equation}
Z\frac{\sin(\beta_ka)}{\beta_ka} + \cos(\beta_ka) = \cos(ka),
\end{equation}
with $Z>0$ being an effective strength of the potential. In the extended zone scheme,
the energies $E(k)$ are discontinuous functions for $k=n\pi/a$,
where $n=\ldots-1,0,1\ldots$. In this convention, if $n\pi/a \leq k \leq (n+1)\pi/a$,
the energies $E(k)$ belong to the $n$-th band and the Bloch states are characterized
by {\it one} quantum number $k$. Because $A_k(z)$ is a periodic function,
one may expand it in the Fourier series $ A_k(z) = \sum_n A_n \exp(ik_nz)$, where $k_n=2\pi n /a$.

In the Heisenberg picture the time-dependent velocity averaged over a wave packet $f(z)$ is
\begin{equation} \label{Kron_v0}
\langle \hat{v}(t)\rangle = \frac{\hbar}{m_0} \int \hspace*{-0.5em}
        \int \!\!dk dk' \langle f|k\rangle\langle k|\frac{\partial}{i\partial z}|k'\rangle
                  \langle k'|f\rangle e^{i(E_k-E_{k'})t/\hbar},
\end{equation}
where $|k\rangle$  is the Bloch state. The matrix elements of momentum
are  $\langle k|\hat{p}|k'\rangle = \hbar \delta_{k',k+k_n}K(k,k')$,
where
\begin{equation}
 K(k,k') = \int_0^a \psi_{k}(z)^*\frac{\partial\psi_{k'}(z)} {i\partial z} dz,
\end{equation}
The wave packet $f(z)$ is taken in a Gaussian form of the width $d$ and centered at $k_0$, and its
matrix elements are $\langle f|k\rangle = \sum_n A_nF(k,k_n)$, where
\begin{equation}
F(k,k_n)=\int_{-\infty}^{\infty} f^*(z)e^{iz(k+k_n)} dz.
\end{equation}
Inserting the above matrix elements to Eq. (\ref{Kron_v0}) we obtain
\begin{eqnarray} \label{Kron_v1}
\langle \hat{v}(t)\rangle = \frac{\hbar}{m_0}\sum_{n,n',l}\int \hspace*{-0.5em} \int \!\!dk dk'
        A^*_{n}A_{n'} F^*(k,k_{n}) F(k',k_{n'}) \times \nonumber \\
  K(k,k') e^{i(E_k-E_{k'})t/\hbar} \delta_{k',k+k_l}.
\end{eqnarray}

Figure 2 shows results for the electron ZB, as computed for a superlattice.
The electron velocity and position are indicated. As follows from the inset in Fig. 3,
a relative narrowness of the wave
packet in $k$ space cuts down contributions from $k$ values away from $k_0$,
and one deals effectively with the  vicinity of {\it one} energy gap.
It is seen that for a superlattice with the period $a=200$\AA\ the ZB displacement is
about $\pm 50$\AA, i.e. a fraction of the period, in
agrement with the rough estimations given above. The period of oscillations is of the order of several
picoseconds.

\begin{figure}
\includegraphics[width=8.5cm,height=8.5cm]{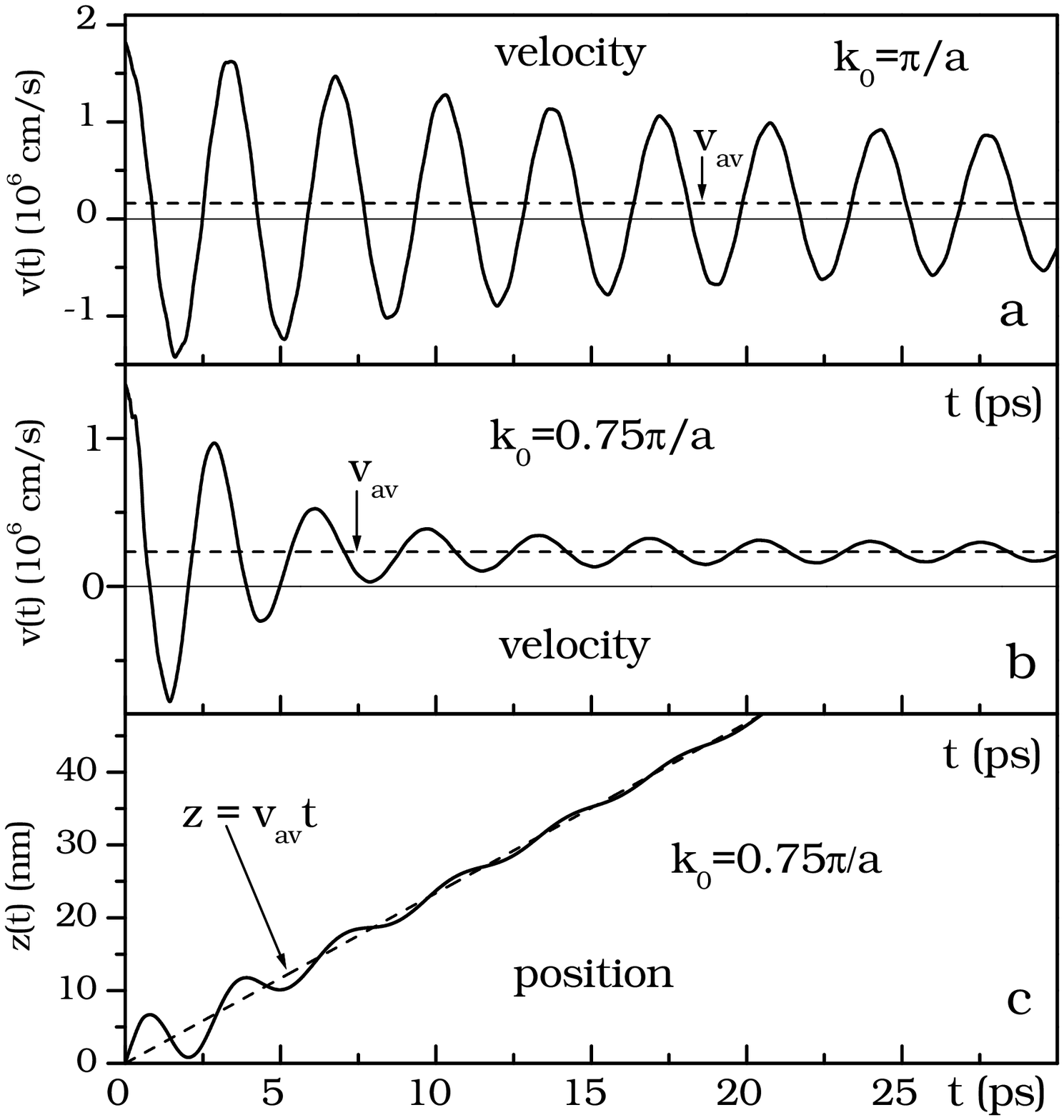} \label{Fig2}
\caption{Calculated electron ZB velocities and position in a superlattice versus time.
         The packet width is $d=400$\AA, Kronig-Penney parameter is $Z=1.5\pi$,
         superlattice period is $a=200$\AA. (a) Packet centered at $k_0=\pi/a$;
         (b) and (c) packet centered at $k_0=0.75\pi/a$.
         The dashed lines indicate motions with average velocities.}
\end{figure}

\begin{figure}
\includegraphics[width=8.5cm,height=8.5cm]{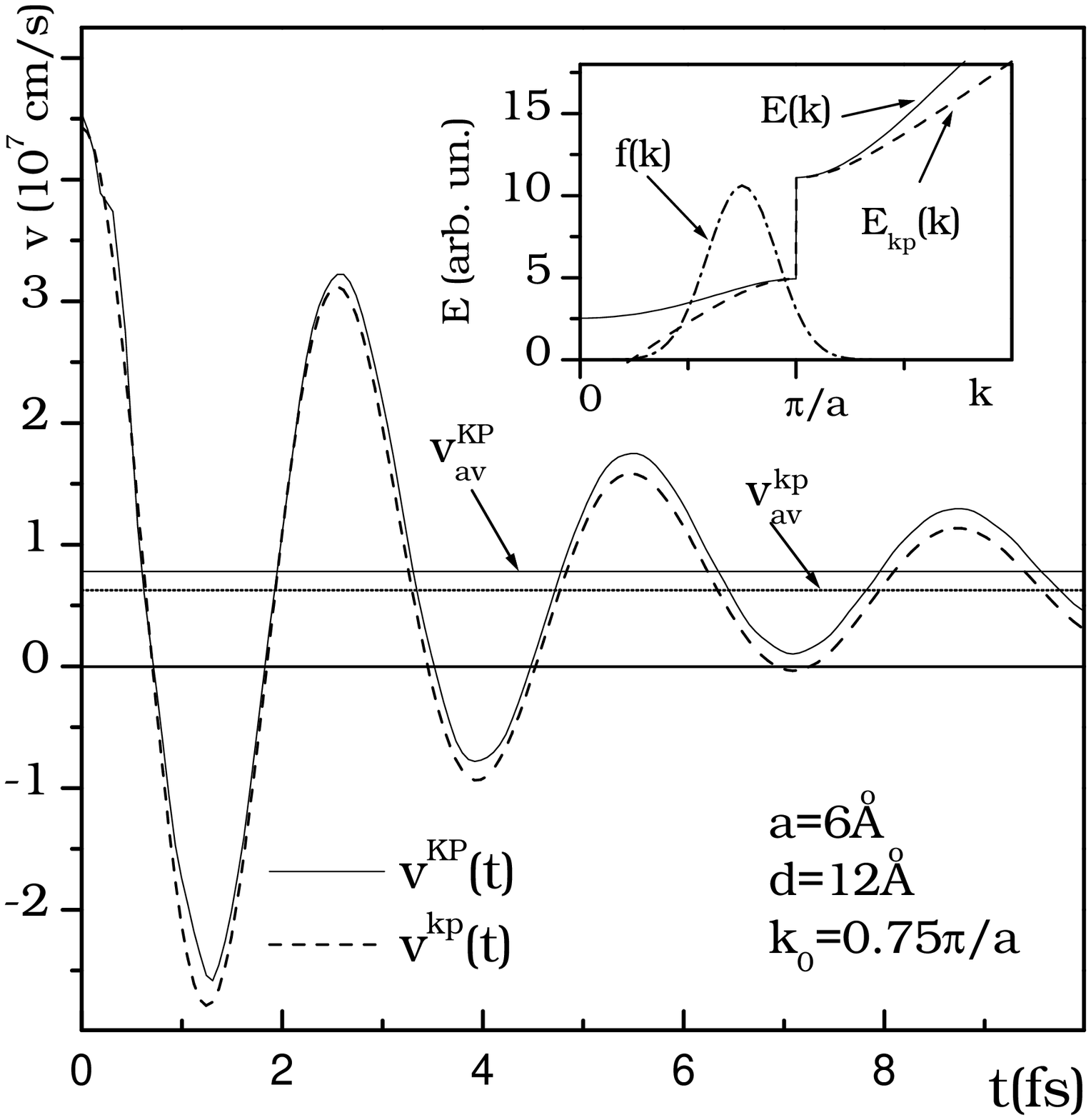}
\caption{ZB of electron velocity in a periodic lattice versus time. Solid line: the complete Kronig-Penney
         model, dashed line: the two-band {\bf k.p} model. Inset: Calculated bands for the Kronig-Penney
         (solid line) and the two-level {\bf k.p} model (dashed line) in the vicinity of $k=\pi/a$.
         The wave packet $f(k)$ centered at $k_0=0.75\pi/a$ is also indicated (not normalized).}
\label{Fig3}
\end{figure}

Finally, we want to demonstrate that the two-band {\bf k.p} model, used until present to calculate the
Zitterbewegung \cite{Rusin09}, is adequate for a description of this phenomenon. We calculate the packet
velocity near the point $k_0=\pi/a$ for a one-dimensional Kronig-Penney periodic Hamiltonian using
the Luttinger-Kohn (LK) representation \cite{Luttinger55}. The LK functions $\chi_{nk}(z)=e^{ikz}u_{nk_0}(z)$,
where $u_{nk_0}(z)=u_{nk_0}(z+a)$, also form a complete set. We have
\begin{equation}
\langle \hat{v}(t)\rangle = \langle f|\hat{v}(t)|f\rangle =
\sum_{kk'nn'}\langle f|nk\rangle\langle nk|\hat{v}(t)|n'k'\rangle\langle\ n'k'|f\rangle,
\end{equation}
where the velocity in the Heisenberg picture is
$\hat{v}(t)=(\hbar/m_0)e^{i\hat{H}t/\hbar}(\partial /i\partial z)e^{-i\hat{H}t/\hbar}$. Restricting
the above summation to the conduction and valence bands: $n,n'=1,2$, we obtain in the matrix form
\begin{equation} \label{kp_vt}
\langle \hat{v}(t)\rangle \approx  \left(\begin{array}{c} f_1 \\ f_2 \end{array}\right)^{\dagger}
 \left(\begin{array}{cc} \hat{v}(t)_{11}& \hat{v}(t)_{12} \\ \hat{v}(t)_{21}& \hat{v}(t)_{22} \end{array}\right)
 \left(\begin{array}{c} f_1 \\ f_2 \end{array}\right),
\end{equation}
where $f_i=\langle nk|f\rangle$, and $\hat{v}(t)_{nn'}$ are the
matrix elements of the time-dependent velocity operator between the LK functions.
Equation (\ref{kp_vt}) looks like the {\bf k.p} approach to ZB used previously.

The {\bf k.p} Hamiltonian is obtained from the initial periodic Hamiltonian in the standard way
\cite{Luttinger55,Kane57}. In the two-band model the result is
\begin{equation} \label{kp_Hkp}
 \hat{H}_{kp}= \left(\begin{array}{cc} \hbar^2k^2/2m_0 + E_1 & \hbar kp_{21}/m_0 \\
       \hbar kp_{12}/m_0 & \hbar^2k^2/2m_0 + E_2 \end{array}\right),
 \end{equation}
where $p_{12}=p_{21}^*$  are the interband elements of momentum, and $E_1$ and $E_2$
are the band-edge energies. The velocity matrix at $t=0$ is $\hat{v}_{kp}=\partial \hat{H}_{kp}/\hbar\partial k$.
The calculation of velocity in the Heisenberg picture is described in Ref. \cite{Rusin07a}.

In Fig. 3 we compare the ZB oscillations of velocity calculated using: (a) real $E(k)$ dispersions
resulting from the Kronig-Penney model and the corresponding Bloch functions of Eq. (\ref{Kron_Ak});
(b) two-band $E(k)$ dispersions obtained from Eq. (\ref{kp_Hkp}) and the corresponding LK functions. It is
seen that, although we take the packet not centered at $k=\pi/a$, the two-band {\bf k.p} model
gives an excellent description of ZB. For $k_0=\pi/a$ the two descriptions are almost identical.
It is seen from Fig. 3 that for a wave packet of the width  $d=2a$ the ZB is already well described
by the {\bf k.p} model.
On the other hand, for a much wider packet in $k$ space the two descriptions differ
more, especially when the $k$-width encompasses more than one energy gap.

A few remarks are in order. The transient character of ZB, illustrated in Figs. 2 and 3,
is a result of describing the electron dynamics with the use of wave packets, see \cite{Lock79}.
In particular, it is seen that wider packets (in real space) result in longer transient times.
In the limiting cases of plane waves the electron oscillates indefinitely, similarly to
the classical description \cite{Zawadzki05KP}. Second, one should bear in mind that the
standard conductivity theories use {\it average} electron velocities $\bar{v}=\hbar k/m^*$, as
indicated in the above figures. Our work shows that at very short times the electron dynamics is
completely different from its average behavior. Third, both periods and amplitudes of ZB, as
shown in Fig. 2, are comparable to those appearing in the Bloch
oscillation measurements \cite{Lyssenko96}, so the ZB should be also observable experimentally.
Clearly, it is difficult to follow the behavior of a single electron and, in order to observe
the trembling motion, one should produce many electrons moving in phase. This can
be  most readily done using laser pulses, see \cite{Rusin09,Lyssenko96}.

In summary, we considered fundamentals of electron motion in periodic structures and
showed that the extensively studied phenomenon of electron Zitterbewegung in crystalline solids
is caused by oscillations of velocity assuring the total energy conservation as an electron moves in
a periodic potential. This means that, although the ZB in solids was often studied in literature
using the two-band {\bf k.p} model of band structure analogous to the Dirac equation for relativistic electrons
in a vacuum, the origins of ZB in s solid and in a vacuum are completely different. We also performed a
rigorous quantum calculation of ZB for an electron in the Kronig-Penney potential and showed that the
two-band {\bf k.p} model is adequate for its description.

We dedicate this work to the memory of Professor R. A. Smith, whose excellent book ''Wave Mechanics of
Crystalline Solids'' was very helpful in our endeavor.
This work was supported in part by The Polish Ministry of Science and Higher Education through Laboratory of
Physical Foundations of Information Processing.

\end{document}